\documentclass[aps,prb,showpacs,reprint,superscriptaddress]{revtex4-1}

\usepackage{graphicx}
\usepackage{color}
\usepackage{amsmath}
\usepackage{bbm}
\usepackage{amssymb}

\usepackage{dsfont}

\usepackage[utf8]{inputenc}	
\usepackage[T1]{fontenc}

\usepackage[svgnames]{xcolor}

\input epsf

\begin{document}

\title{Topological superconductivity with mixed singlet-triplet pairing in moir\'e transition-metal-dichalcogenide bilayers}
\author{Waseem Akbar}
\affiliation{Academic Centre for Materials and Nanotechnology, AGH University of Krakow, Al. Mickiewicza 30, 30-059 Krakow,
Poland}
\author{Andrzej Biborski}
\affiliation{Academic Centre for Materials and Nanotechnology, AGH University of Krakow, Al. Mickiewicza 30, 30-059 Krakow,
Poland}
\author{Louk Rademaker}
\affiliation{Department of Quantum Matter Physics, University of Geneva, CH-1211 Geneva, Switzerland}
\author{Micha{\l} Zegrodnik}
\email{michal.zegrodnik@agh.edu.pl}
\affiliation{Academic Centre for Materials and Nanotechnology, AGH University of Krakow, Al. Mickiewicza 30, 30-059 Krakow,
Poland}

\begin{abstract}
We investigate strong coupling topological superconductivity in twisted moiré bilayer WSe$_2$. Our approach is based on an effective $t$-$J$ model with displacement-field-dependent complex hoppings, which is treated with the variational Gutzwiller projection method. The calculated phase diagram contains domes of topologically nontrivial superconducting phases, with Chern numbers $C=\pm 2,\;\pm 4$. The order parameter is characterized by a mixed $d$+$id$-wave (singlet) and $p$-$ip$-wave (triplet) gap symmetry. We also report on the appearance of an additional topologically trivial extended $s$-wave and $f$-wave paired phase. 
As we show, by changing the electron density and displacement field, one can tune the singlet and triplet contributions to the pairing, as well as induce topological phase transitions between superconducting states characterized by different values of the Chern number.
We analyze the physical origin of the reported effects and discuss it briefly in the view of new possibilities for designing unconventional superconductivity in moir\'e systems.
\end{abstract}

\maketitle

\section{Introduction}

In recent years, more and more two-dimensional systems realizing a moir\'e pattern of atoms have been investigated, as they provide a fertile ground for novel electronic states. The characteristic moir\'e structure emerges after twisting two or more monolayers with respect to each other, like in the well-known twisted bilayer graphene\cite{Cao2018_1, Cao2018_2,Balents.2020}. A similar effect can also be obtained as a result of lattice mismatch in heterostructures without the necessity of rotational misalignment\cite{Tang2020, Li2021}. 

The common feature of moiré systems is the appearance of flat electronic bands, which implies a significant role of electronic interactions. For graphene-based multilayers, the flat bands can be realized only for specific twist angles (so-called magic angles), for which the system exhibits an unusually rich phase diagram\cite{Cao2018_1, Cao2018_2,Balents.2020}. For transition metal dichalcogenide (TMD) moir\'e structures, the appearance of flat bands is more robust with respect to the twist angle deviations\cite{Wang2020,Ghiotto2021,Li2021,Rademaker_2023_flat_bands,Kennes.2021,Mak.2022}. Evidence of strongly correlated phenomena in the moir\'e TMDs has been reported in recent years: amongst them metal-insulator transitions\cite{Ghiotto2021,Wang2020}, different forms of charge ordering\cite{Jin2021,Cui2021}, as well as signatures of superconductivity\cite{Wang2020,An2020}. We write `signatures' because the supposed superconducting state has not been unambiguously detected experimentally. Nevertheless, what distinguishes TMDs from graphene-based moiré systems is that the former have a relatively strong spin-orbit coupling. This would naturally imply that the superconducting state in the TMDs can have some nontrivial pairing symmetry.

The main focus of this paper is a theoretical exploration of the possible superconducting phases in a twisted WSe$_2$ homobilayer (tWSe$_2$), for which `signatures' of superconductivity were observed.\cite{Wang2020} 
The low energy physics of tWSe$_2$ is believed to be described by a single band Hamiltonian on a triangular lattice with both spin- and direction-dependent complex hoppings\cite{Wang2020,Haining2020}. The Hubbard $U$ resulting from such an approach is approximately $U\gtrsim W$ ($W$ is the bare band width), which suggests that tWSe$_2$ is in a relatively strongly correlated regime\cite{Wang2020, Haining2020, Rademaker2023_Wigner_Mott}. 
Indeed, the experimental phase diagram suggests an insulating state at half-filling with adjacent superconducting domes. Even though the pairing mechanism and the symmetry of the superconducting gap are still unknown, the dome structure implies that the superconductivity might arise purely from strong electron repulsion through Hubbard or $t$-$J$ models\cite{Belanger2022_cDMFT_SC,Sheng2023}. 
The Hubbard model as applied to the description of homo- and heterobilayer moir\'e TMDs has been analyzed in the view of spin- and charge-ordering, as well as the appearance of unconventional superconducting state\cite{Zang2022,Rademaker2023_Wigner_Mott,Rademaker_2023_Kagome,Wu_2023_PDW_SC,Belanger2022_cDMFT_SC, Zang2022, biborski2024}. 
For the case of the paired phase these approaches resemble those that have been applied earlier to the single-band models of the well-known cuprates\cite{Spalek2022}. However, other proposals related to TMD superconductivity, based on the spin-valley fluctuations or interlayer excitonic physics, have also been put forward\cite{schrade2021nematic,Millis2023}.

In this manuscript we study the $t$-$J$ model relevant for tWSe$_2$ using the Gutzwiller approximation, and analyze in detail the topological features of the obtained unconventional paired states.
The interesting peculiarity about TMD systems is that the moiré pattern generates a triangular lattice system, which allows for quite nontrivial and possible topological order parameter symmetries. 
In fact, we predict a two-dome structure of {\em mixed singlet-triplet superconductivity} (with $d+id/p-ip$ gap symmetry mixing), separated by a correlation-induced insulating state at half filling. This is consistent with our previous work on the $t$-$J$-$U$ model as applied to the description of tWSe$_2$\cite{Zegrodnik2023}.
Additionally, here we show that the obtained mixed state is topologically nontrivial with Chern number values $C=\pm 2,\;\pm 4$. 
Furthermore, we focus on the effect of the displacement field by explicitly taking into account the appropriate complex hopping parameters. 
By tuning the displacement field one can induce topological phase transitions between phases with different Chern numbers, as well as control the balance between the singlet and triplet contributions to the pairing. Additionally, for low and high electron concentration regimes we have obtained a topologically trivial extended $s$-wave/$f$-wave paired state. Our results show the TMD homobilayers as promising candidates for topological superconductors characterized by a high degree of tunability. 

The remainder of this publication is organized as follows. In Sec.~\ref{Sec:Model} we introduce the $t$-$J$ model relevant for tWSe$_2$ and we describe our method of the Gutzwiller projected variational wavefunction. In Sec.~\ref{Sec:ResultsA} we provide our results on the structure of the mixed singlet-triplet superconducting domes, while in Sec.~\ref{Sec:ResultsB} we discuss the topological classification of the phases. Finally, in Sec.~\ref{Sec:Outlook} we provide an outlook on the experimental detection of topologically nontrivial superconductivity in tWSe$_2$.


\section{Model and method}
\label{Sec:Model}
In this work we apply a moir\'e band $t$-$J$ Hamiltonian on a triangular lattice, i.e.,
\begin{equation}
\begin{split}
 \hat{H}=& \sum_{\langle ij\rangle\sigma}t_{ij\sigma}\;\hat{c}^{\dagger}_{i\sigma}\hat{c}_{j\sigma} \\
 +& J\sideset{}{'}\sum_{\langle ij\rangle}\bigg(\hat{S}^z_i\hat{S}^z_j +\cos{(2\phi_{ij\uparrow})}\sum_{\alpha=x,y}\hat{S}^{\alpha}_i\hat{S}^{\alpha}_j\\
    +&\sin{(2\phi_{ij\uparrow})}(\mathbf{\hat{S}}_i\times\mathbf{\hat{S}}_j)\cdot \hat{z} \bigg), 
 \label{eq:Hamiltonian_start}
 \end{split}
\end{equation}
where $\hat{c}^{\dagger}_{i\sigma}$ ($\hat{c}_{i\sigma}$) are the creation (anihilation) operators for electrons at site $i$ with spin $\sigma=\{1,-1\}$, and $\mathbf{\hat{S}}_i=(\hat{S}^x_i,\hat{S}^y_i, \hat{S}^z_i)$ is the spin-$\frac{1}{2}$ operator. The summation over $\langle ij\rangle$ in both terms is restricted to nearest neighbors only since the longer-ranged contributions are expected to be about one order of magnitude smaller. The primed summation in the interaction part means that each bond appears only once. The complex hopping parameters, $t_{ij\sigma}$, fulfill the hermiticity requirement ($t_{ij\sigma}=t_{ij\sigma}^{*}$), time-reversal symmetry condition ($t_{ij\sigma}=t_{ij\bar{\sigma}}^{*}$), threefold rotational symmetry ($C_3$), and can be explicitly expressed in the following form
\begin{equation}
    t_{ij\sigma}=|t|e^{i\phi_{ij\sigma}}=|t|e^{i\sigma\nu_{ij}\phi},
    \label{eq:hoppings_complex}
\end{equation}
with $\nu_{ij}=\pm 1$, depending on the direction of the bond. The resulting complex phases of the hoppings for all the six nearest neighbors are shown schematically in Fig. 1. As one can see from Eq. (\ref{eq:Hamiltonian_start}) the exchange interaction $\sim J$ is supplemented with the Dzyaloshinskii-Moriya term which results from the fact that the phase of the hoppings are spin-dependent. 

This form of the $t$-$J$ model can be derived from the Hubbard model as applied to the description of the tWSe$_2$ in the large $U$ limit. As shown in Refs. \onlinecite{Wang2020,Haining2020}, within such approach, both the absolute values of the hoppings ($|t|$) as well as their complex phase ($\phi$) depend on the bias voltage across the bilayer (the so-called displacement field $D$). This provides an experimentally controllable parameter which can be used to tune $in$ $situ$ the magnitude of the valley-dependant spin-splitting as well as the position of the Van Hove singularity (cf. Fig. \ref{fig:FS_and_DOS}). The values of the complex hopping parameters for the range of displacement fields $D\in[0,0.9]$ V/nm, that are used for the analysis provided in Section 2, have been taken from Ref. \onlinecite{Wang2020}. 

\begin{figure}
 \centering
 \includegraphics[width=0.5\textwidth]{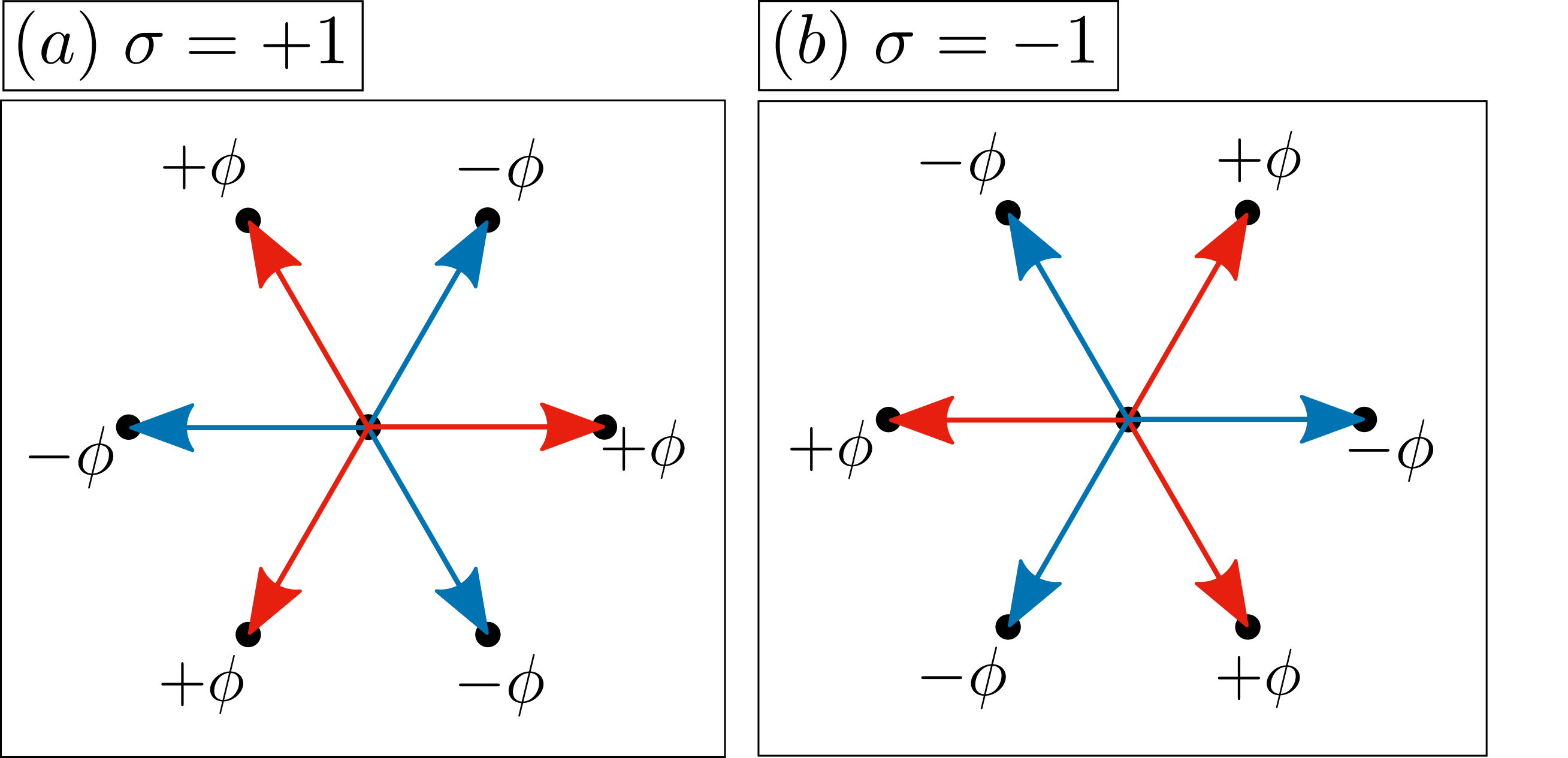}
 \caption{Phase of the complex hoppings to the nearest neighbors for spin up (a) and spin down (b) electrons} 
 \label{fig:hopping_scheme}
\end{figure}

\begin{figure}
 \centering
 \includegraphics[width=0.5\textwidth]{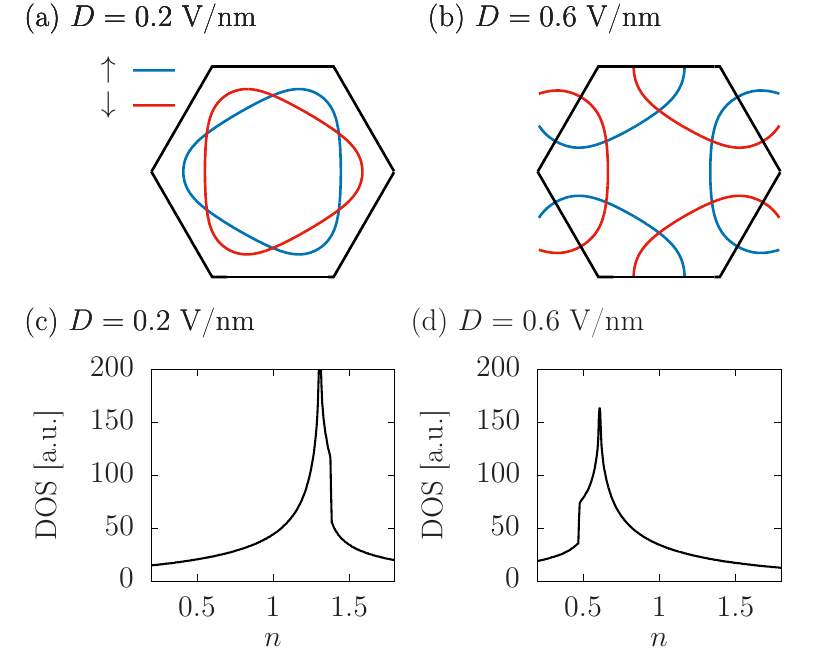}
 \caption{The spin-dependent Fermi surfaces at half filling (a,b) and the density of states (c,d) for two selected values of the displacement field, $D$. The data have been obtained based on the single particle part of Hamiltonian given by Eq. (\ref{eq:Hamiltonian_start}). Note that by changing the displacement field one can tune the valley-dependent spin-splitting as well as the position of the Van Hove singularity. The Fermi surface in (a) and (b) are of electron-like and hole-like character, respectively.} 
 \label{fig:FS_and_DOS}
\end{figure}

The $t$-$J$ model is one of the canonical models used for the description of strongly correlated electron systems. In order to take into account electron-electron correlations, we apply the variational wave function of the Gutzwiller type which is defined in the following manner
\begin{equation}
    |\Psi_G\rangle=\hat{P}|\Psi_0\rangle,
    \label{eq:correlated_state}
\end{equation}
where $|\Psi_0\rangle$ is the non-correlated (mean-field) state and $\hat{P}$ is the correlation operator 
\begin{equation}
    \hat{P}=\prod_i\sum_{\Gamma}\lambda_{i,\Gamma}|\Gamma\rangle_{i\;i}\langle\Gamma|,
\end{equation}
where $i$ runs over all the lattice sites, $|\Gamma\rangle\in \{ |\emptyset\rangle,\;|\uparrow\rangle,\;|\downarrow\rangle,\;|\uparrow\downarrow\rangle \}$, and $\lambda_{i,\Gamma}$ are the corresponding variational parameters. 
It has been shown that for such a form of the correlation operator it is convenient to impose an additional condition on the $\hat{P}_i$ operator\cite{Bunemann_2012}, i.e.,  
\begin{equation}
    \hat{P}^2_i=1+x\hat{d}_i^{HF},
\end{equation}
where $\hat{d}^{HF}_i=\hat{n}^{HF}_{i\uparrow}\hat{n}^{HF}_{i\downarrow}$, $\hat{n}^{HF}_{i\sigma}=\hat{n}_{i\sigma}-n_{i\sigma}$, with $n_{i\sigma}=\langle \Psi_0|\hat{n}_{i\sigma}|\Psi_0\rangle$, and $x$ is yet another variational parameter. It is straightforward to show that
\begin{equation}
\begin{split}
    \lambda^2_{\uparrow\downarrow}&=1+x(1-n_s)^2,\\
    \lambda^2_s&=1-xn_s(1-n_s)^2,\\
    \lambda^2_{\emptyset}&=1+xn_s^2,
\end{split}
\end{equation}
where for simplicity we have assumed a homogeneous system with no magnetic or charge ordering. Hence, $\lambda_{i,\Gamma}\equiv \lambda_{\Gamma}$ and $\lambda_{\uparrow}=\lambda_{\downarrow}=\lambda_{s}$, $n_{i\uparrow}=n_{i\downarrow}\equiv n_{s}$. As a result, we are left with only one variational parameter, $x$. Now, in order to project out the double occupancies, what is required for the $t$-$J$ model, one has to set 
\begin{equation}
\begin{split}
\mathrm{for}\;n_s<0.5:&\;\lambda_{\uparrow\downarrow}=0\;\Rightarrow x=-\frac{1}{(1-n_s)^2},\\
\mathrm{for}\;n_s>0.5:&\;\lambda_{\emptyset}=0\;\Rightarrow x=-\frac{1}{n_s^2},
\end{split}
\end{equation}
meaning that for electron concentrations below half-filling the double occupancies of electrons are prohibited, and above half-filling the holons are prohibited.

In order to obtain the formula for the expectation value of our hamiltonian per lattice site,
\begin{equation}
    E_G=\frac{\langle\Psi_G|\hat{H}|\Psi_G\rangle}{N\langle\Psi_G|\Psi_G\rangle}=\frac{1}{N}\langle\hat{H}\rangle_G,
    \label{eq:ground_state_energy}
\end{equation}
we apply the diagrammatic expansion of the Gutzwiller wave function (DE-GWF)\cite{Spalek2022} in the zeroth order which is equivalent to the {\em Statistically consistent Gutzwiller Approximation} (SGA)\cite{Abram2013}. As a result one obtains
\begin{equation}
\begin{split}
    \langle\hat{H}\rangle_G&=\sum_{ij}q^2t_{ij\sigma}\langle\hat{c}^{\dagger}_{i\sigma}\hat{c}_{j\sigma}\rangle_0,\\
    &+\lambda^4_s\;J\sideset{}{'}\sum_{\langle ij\rangle}\bigg(\frac{1}{2}\;\sum_{\sigma}\;e^{i\sigma2\phi_{ij}}\langle\hat{c}^{\dagger}_{i\sigma}\hat{c}_{i\bar{\sigma}}\hat{c}^{\dagger}_{j\bar{\sigma}}\hat{c}_{j\sigma} \rangle_0\\
    &+\frac{1}{4}\;\sum_{\sigma\sigma'}\sigma\sigma'\langle\hat{n}^{HF}_{i\sigma}\hat{n}^{HF}_{j\sigma'}\rangle_0\bigg),
\end{split}
\label{eq:hamiltonian_expectation_value}
\end{equation}
where $q=\lambda_s\big(\lambda_d n_s+\lambda_{\emptyset}(1-n_s)\big)$ and $\langle\hat{o}\rangle_0$ stands for the expectation value of the $\hat{o}$ operator in the state $|\psi\rangle_0$. By applying the Wick's theorem one can decompose all the four-operator expectation values which appear on the right-hand side of the above equation. Hence, the obtained $E_G$ becomes a function of $x$, $n_s$ as well as the electron hopping and Cooper pairing mean field parameters,
\begin{equation}
\begin{split}
    P_{ij\sigma}=\langle\hat{c}^{\dagger}_{i\sigma}\hat{c}_{j\sigma}\rangle_0,\quad S_{ij}^{\sigma\sigma'}=\langle\hat{c}_{i\sigma}\hat{c}_{j\sigma'}&\rangle_0,\\
\end{split}
\label{eq:mean_field_params}
\end{equation}
respectively. It should be noted that the expectation values in the non-correlated state together with the variational parameters determine the corresponding expectation values in the correlated state. Namely,
\begin{equation}
\begin{split}
    \Delta^{\sigma\sigma'}_{ij}&=\langle\hat{c}_{i\sigma}\hat{c}_{j\bar{\sigma}}\rangle_G=q^2\langle\hat{c}_{i\sigma}\hat{c}_{j\sigma'}\rangle_0,\\
    \Lambda_{ij\sigma}&=\langle\hat{c}^{\dagger}_{i\sigma}\hat{c}_{j\sigma}\rangle_G=q^2\langle\hat{c}^{\dagger}_{i\sigma}\hat{c}_{j\sigma}\rangle_0.
\end{split}
\label{eq:mean_field_corr}
\end{equation}
Within our analysis, we treat $S_{ij}^{\uparrow\downarrow}$ and $S_{ij}^{\downarrow\uparrow}$ separately allowing for both pure spin-singlet pairing ($S_{ij}^{\uparrow\downarrow}=-S_{ij}^{\downarrow\uparrow}$), pure spin-triplet pairing ($S_{ij}^{\uparrow\downarrow}=S_{ij}^{\downarrow\uparrow}$) as well as their mixture ($|S_{ij}^{\uparrow\downarrow}|\neq |S_{ij}^{\downarrow\uparrow}|$). At the same time, we neglect the possibility of $\uparrow\uparrow$ and $\downarrow\downarrow$ pairing due to the fact that here the pairing mechanism is based on the kinetic exchange term which is of antiferromagnetic type. Additionally the $S^{z}=\pm 1$ pairing channels would lead to a Fermi wave vector mismatch, which is detrimental when it comes to the formation of the superconducting state. 

In order to determine the values of the mean fields we apply the Effective Hamiltonian Scheme\cite{Kaczmarczyk2015}. Within such approach the minimization condition of the ground state energy (\ref{eq:ground_state_energy}) leads to an effective Hamiltonian, which for our case has the following form
\begin{equation}
\begin{split}
 \hat{\mathcal{H}}_{\textrm{eff}}&=\sum_{ij\sigma}\tilde{t}_{ij\sigma}\hat{c}^{\dagger}_{i\sigma}\hat{c}_{j\sigma}-\tilde{\mu}\sum_{i\sigma}\hat{n}_{i\sigma}\\
 &+\sideset{}{'}\sum_{\langle ij\rangle\sigma}\big((\tilde{\Delta}_{ij\sigma\bar{\sigma}})^*\hat{c}_{j\sigma}\hat{c}_{i\bar{\sigma}}+h.c.\big),
 \end{split}
 \label{eq:H_effective}
\end{equation}
where the effective hopping, effective chemical potential, and the effective superconducting gap parameters are defined through the corresponding relations
\begin{equation}
 \tilde{t}_{ij\sigma}\equiv \frac{\partial\mathcal{F}}{\partial P_{ij\sigma}},\quad (\tilde{\Delta}_{ij\sigma\bar{\sigma}})^*\equiv \frac{\partial\mathcal{F}}{\partial S^{\sigma\bar{\sigma}}_{ji}},\quad\tilde{\mu}\equiv-\frac{\partial\mathcal{F}}{\partial n_s},
 \label{eq:effective_param}
\end{equation}
where $\mathcal{F}=E_G-2\mu_Gn_s$ with $\mu_G$ being the chemical potential determined in the correlated state. For the sake of completeness we show explicitly the form of the effective model parameters below
\begin{equation}
    \tilde{t}_{ij\sigma}=q^2\;t_{ij\sigma}-J\;\frac{\lambda_s^4}{2}\;\big[2(P_{ij\bar{\sigma}})^*\;e^{i2\phi_{ij}^{\sigma}}+(P_{ij\sigma})^*\big]
\end{equation}

\begin{equation}
   (\tilde{\Delta}_{ij}^{\sigma\bar{\sigma}})^*=-J\;\frac{\lambda_s^4}{2}\; \big[2(S_{ij}^{\sigma\bar{\sigma}})^*\;e^{i2\phi_{ij}^{\sigma}}-(S_{ij}^{\bar{\sigma}\sigma})^*\big]
\end{equation}

\begin{equation}
    \tilde{\mu}=\mu_G-U\lambda^2_d n_s
\end{equation}

After the transformation to the reciprocal space we write down the effective Hamiltonian in the matrix form
\begin{equation}
\begin{split}
\hat{\mathcal{H}}_{\textrm{eff}}&=\sum_{\mathbf{k}\sigma}\begin{array}{c}
 \left(\hat{c}^{\dagger}_{\mathbf{k}\sigma}\;\; \hat{c}_{-\mathbf{k}\bar{\sigma}} \right)\\
 \\
\end{array}\left(\begin{array}{cc}
 \tilde{\epsilon}_{\mathbf{k}\sigma}-\tilde{\mu} & \tilde{\Delta}_{\mathbf{k\bar{\sigma}\sigma}}\\
(\tilde{\Delta}_{\mathbf{k\bar{\sigma}\sigma}})^{*} & -\tilde{\epsilon}_{\mathbf{k}\sigma}+\tilde{\mu} \\
\end{array} \right)\left(\begin{array}{c}
 \hat{c}_{\mathbf{k}\sigma}\\
 \hat{c}^{\dagger}_{-\mathbf{k}\bar{\sigma}}\\
\end{array} \right)\;\\
&+\sum_{\mathbf{k}\sigma}(\tilde{\epsilon}_{\mathbf{k}\sigma}-\tilde{\mu}),
\end{split}
\label{eq:H_matrix}
\end{equation}
where $\bar{\sigma}=-\sigma$ and we have used the relation $\tilde{\epsilon}_{\mathbf{k}\sigma}=\tilde{\epsilon}_{-\mathbf{k}\bar{\sigma}}$ (time reversal symmetry condition). The effective dispersion relations as well as the superconducting gaps in $\mathbf{k}$-space are related with their real-space counterparts in the following manner
\begin{equation}
\begin{split}
    \tilde{\epsilon}_{\mathbf{k}\sigma}&=\sum_{i(j)}\tilde{t}_{ij}e^{ik(\mathbf{R}_i-\mathbf{R}_j)}\\
    \tilde{\Delta}_{\mathbf{k}\bar{\sigma}\sigma}&=\sum_{i(j)}\tilde{\Delta}_{ji\sigma\bar{\sigma}}e^{ik(\mathbf{R}_j-\mathbf{R}_i)},
\end{split}
\label{eq:disp_rel_and_SC_gap_k_space}
\end{equation}
where $i$ index runs over the nearest neighbor lattice sites of site $j$. The $\mathbf{R}_i$ and $\mathbf{R}_j$ vectors point to the $i$ and $j$ lattice sites, respectively. The left-hand sites of the equations above do not depend on $j$ since we assume spaciously homogeneous state. 

The eigenvalues of the hamitlonian matrix in Eq. (\ref{eq:H_matrix}) are the following
\begin{equation}
    \begin{split}
        \lambda_{\mathbf{k}\sigma}=\pm\sqrt{(\tilde{\epsilon}_{\mathbf{k}\sigma}-\tilde{\mu})^2+|\tilde{\Delta}_{\mathbf{k}\bar{\sigma}\sigma}|^2}.
    \end{split}
    \label{eq:eigenvalues}
\end{equation}
From the above equation, one can see that in the SC state the gap $\tilde{\Delta}_{\mathbf{k}\downarrow\uparrow}$ ($\tilde{\Delta}_{\mathbf{k}\uparrow\downarrow}$) opens up at the spin up (spin down) Fermi surface. Moreover, the $\tilde{\Delta}_{\mathbf{k}\downarrow\uparrow}$ and $\tilde{\Delta}_{\mathbf{k}\uparrow\downarrow}$ SC gaps can be transformed into the singlet and triplet components,
\begin{equation}
    \begin{split}
        \tilde{\Delta}^s_{\mathbf{k}}=\frac{1}{\sqrt{2}}\big(\tilde{\Delta}_{\mathbf{k}\uparrow\downarrow}-\tilde{\Delta}_{\mathbf{k}\downarrow\uparrow}\big),\\
        \tilde{\Delta}^t_{\mathbf{k}}=\frac{1}{\sqrt{2}}\big(\tilde{\Delta}_{\mathbf{k}\uparrow\downarrow}+\tilde{\Delta}_{\mathbf{k}\downarrow\uparrow}\big),\\
    \end{split}
    \label{eq:SC_gaps_singlet_triplet_k_space}
\end{equation}
which we can use to rewrite the expressions for the eigenvalues
\begin{equation}
    \begin{split}
        \lambda_{\mathbf{k}\sigma}=\pm\sqrt{(\tilde{\epsilon}_{\mathbf{k}\sigma}-\tilde{\mu})^2+\frac{1}{2}|\tilde{\Delta}^t_{\mathbf{k}}-\sigma\tilde{\Delta}^s_{\mathbf{k}}|^2}.
    \end{split}
    \label{eq:eigenvalues_singlet_triplet}
\end{equation}
The paired state with $\Delta^s_{\mathbf{k}}\neq 0$ and $\Delta^t_{\mathbf{k}} \neq 0$ is referred to here as the {\em mixed singlet-triplet} paired state.

The self-consistent equations for the mean field parameters  [Eq.~(\ref{eq:mean_field_params})] can be derived in a standard manner by applying the Bogolubov-de Gennes approach to the effective Hamiltonian given by Eq.~(\ref{eq:H_matrix}). By solving the set of self-consistent equations numerically we determine the hopping and pairing amplitudes to all the six nearest neighbors. Then, in order to obtain their correlated state counterparts, we use Eqs.~(\ref{eq:mean_field_corr}). Similarly as in our previous study\cite{Zegrodnik2023}, to identify the symmetry of the resultant paired state, we transform the six real-space nearest neighbor pairing amplitudes into six symmetry resolved pairing amplitudes of the following form
\begin{equation}
    \Delta^{\sigma\bar{\sigma}}_{M,p}=\frac{i^p}{6}\sum_{i(j)} e^{-iM\theta_{ji}}\Delta^{\sigma\bar{\sigma}}_{ji},
    \label{eq:gap_symmetry_resolved}
\end{equation}
where the summation runs over the six nearest neighbor lattice sites $j$ surrounding site $i$, and $\theta_{ji}$ are the angles $\{0,\;\pi/3,\;2\pi/3,\;\pi,\;4\pi/3,\;5\pi/3\}$ between the positive half-$x$ axis and the $\mathbf{R}_{ji}=\mathbf{R}_j-\mathbf{R}_i$ vector. $M$ is the symmetry factor, which takes integer values and corresponds to six possible pairing symmetries (cf. Table \ref{tab:symmetries}). The $p$ parameter corresponds to the parity of the particular symmetry: for even parity symmetries we have $p=0$, and for odd parities we have $p=1$.
\begin{table}

\begin{center}
\begin{tabular}{ c|c|c|c|c } 
 \hline\hline
 M & p & gap symmetry & parity & spin state \\ 
 \hline
 0 & 0 & extended $s$ & even & singlet \\ 
 \hline
 1 & 1  & $p_x+i\;p_y$ & odd & triplet \\ 
 \hline
 2 & 0  & $d_{x^2-y^2}+i\;d_{xy}$ & even & singlet \\ 
 \hline
 3 & 1  & $f$ & odd & triplet \\ 
 \hline
 4 & 0  & $d_{x^2-y^2}-i\;d_{xy}$ & even & singlet \\ 
 \hline
 5 & 1  & $p_x-i\;p_y$ & odd & triplet \\ 
 \hline\hline
 \end{tabular}
 \label{tab:symmetries}
\end{center}
 \caption{The six possible values of the symmetry factor, $M$, and the corresponding values of the parity factor $p$ (second column), appearing in Eq. (\ref{eq:gap_symmetry_resolved}). The corresponding six symmetries of the superconducting gap together with their parities are provided in the third and fourth columns. The Cooper pair spin state compatible with a given symmetry is provided in the fifth column.}
\end{table}

Finally, similarly as for the case of the momentum space [Eqs.~\ref{eq:SC_gaps_singlet_triplet_k_space}] one can write down the expressions for the real-space singlet and triplet gap amplitudes in the correlated state, which will play the central role in our subsequent analysis
\begin{equation}
\begin{split}
    \Delta^{s}_{M,p}&=(\Delta^{\uparrow\downarrow}_{M,p}-\Delta^{\downarrow\uparrow}_{M,p})/\sqrt{2},\\
    \Delta^{t}_{M,p}&=(\Delta^{\uparrow\downarrow}_{M,p}+\Delta^{\downarrow\uparrow}_{M,p})/\sqrt{2}.\\
    \label{eq:gap_symmetries_spin}
\end{split}
\end{equation}
For the sake of clarity, in the following Sections, we use the symmetry names ($p\pm ip$, $d\pm id$, $f$ etc.) in the subscripts of the symmetry resolved superconducting gaps instead of the values of the $M$ and $p$ factors.


\section{Results}

\subsection{General features of the paired state}
\label{Sec:ResultsA}
We now characterize the unconventional superconducting state within the $t$-$J$ model description of the WSe$_2$ homobilayer at twist angle $\theta = 5.08^\circ$. This is the twist angle where, in Ref.~\onlinecite{Wang2020}, `signatures of superconductivity' have been experimentally observed. In particular, they observe an insulating state at half-filling ($n=1$ in our notation), with possible superconducting domes around $n \approx 0.86$ and $n\approx 1.16$. Note, that in Ref.~\onlinecite{Wang2020} the band filling is expressed as hole density per spin, in contrast to our paper where we use the electron density per lattice site.

\begin{figure}[!t]
 \centering
 \includegraphics[width=0.5\textwidth]{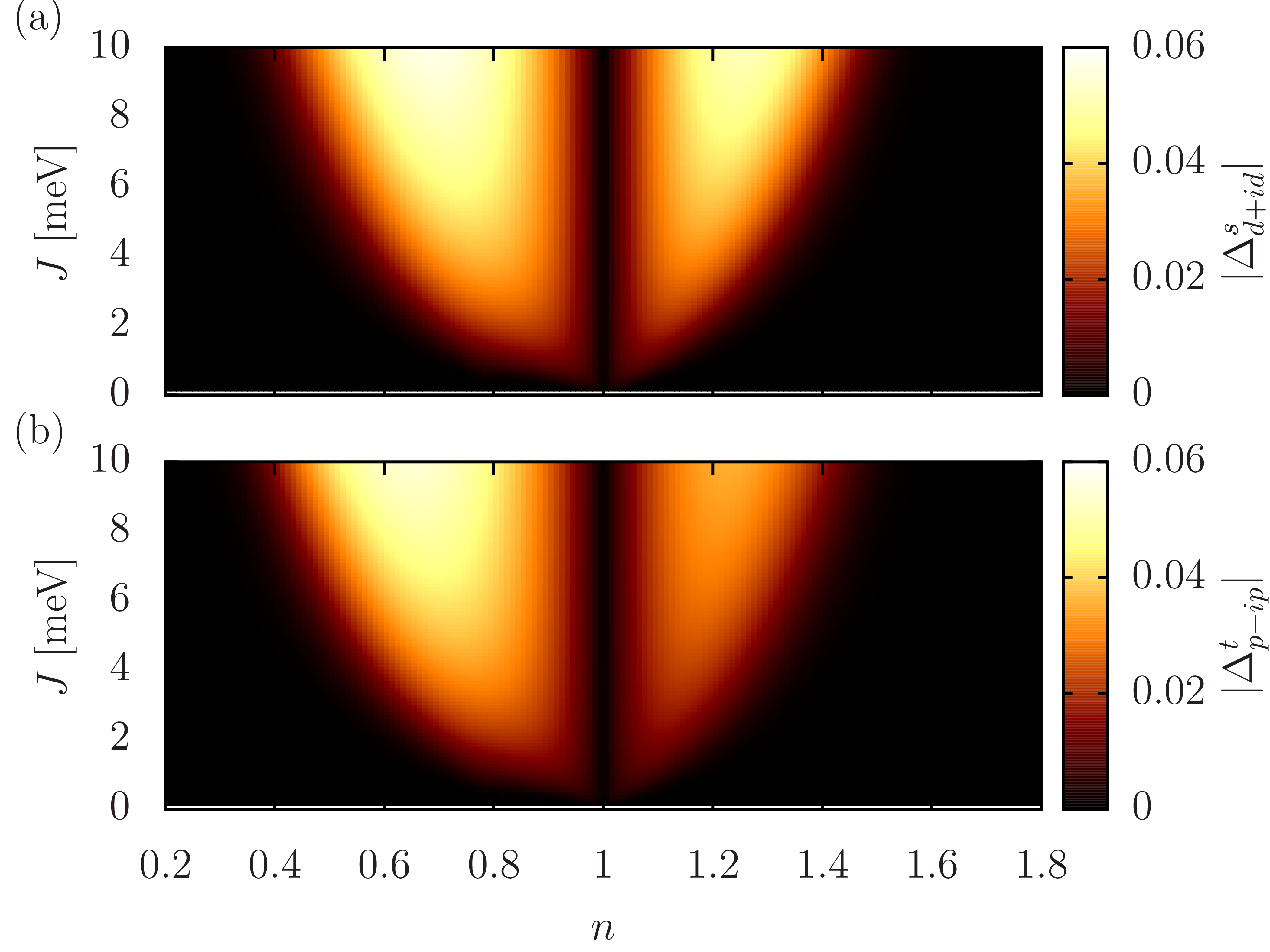}
 \caption{Superconducting gap amplitudes for the spin-singlet $d+id$ (a) and spin-triplet $p-ip$ (b) symmetries as a function of band filling $n$ and exchange interaction energy $J$ for the selected value of the displacement field $D=0.4$\;V/nm.} 
 \label{fig:Delt_nJdep}
\end{figure}

First, we set the hopping parameters to those corresponding to the displacement field $D=0.4$\;V/nm and determine the stability range and the symmetry of the obtained superconducting state. We scan the whole $(n,J)$-plane, where $n=2n_s$ is the total number of particles per lattice site, which we refer to as band filling. As one can see from Fig. \ref{fig:Delt_nJdep} a two-dome structure (similar to the experimental results\cite{Wang2020}) of an unconventional paired state appears, which is characterized by nonzero $d+id$ spin-singlet as well as $p-ip$ spin-triplet amplitudes. Both gap amplitudes show very similar behavior, though the triplet component is slightly smaller. It should be noted that these are not two separate solutions with different gap symmetries, but a single solution for which both amplitudes are nonzero (a mixed singlet-triplet state) meaning that $|\tilde{\Delta}_{\mathbf{k}\uparrow\downarrow}|\neq|\tilde{\Delta}_{\mathbf{k}\downarrow\uparrow}|$ [compare to Eqs. (\ref{eq:eigenvalues}) and (\ref{eq:SC_gaps_singlet_triplet_k_space})]. The appearance of a mixed paired state is a consequence of the valley-dependent spin splitting. Indeed, we verified that if one would have purely real hoppings without spin-orbit coupling, there would be a doubly degenerate Fermi surface with a pure spin-singlet $d+id$ solution. Finally, since the exchange interaction term $\sim J$ is responsible for the appearance of the pairing, the band filling range for which the SC state is stable is wider for larger values of $J$.

\begin{figure}[!b]
 \centering
 \includegraphics[width=0.49\textwidth]{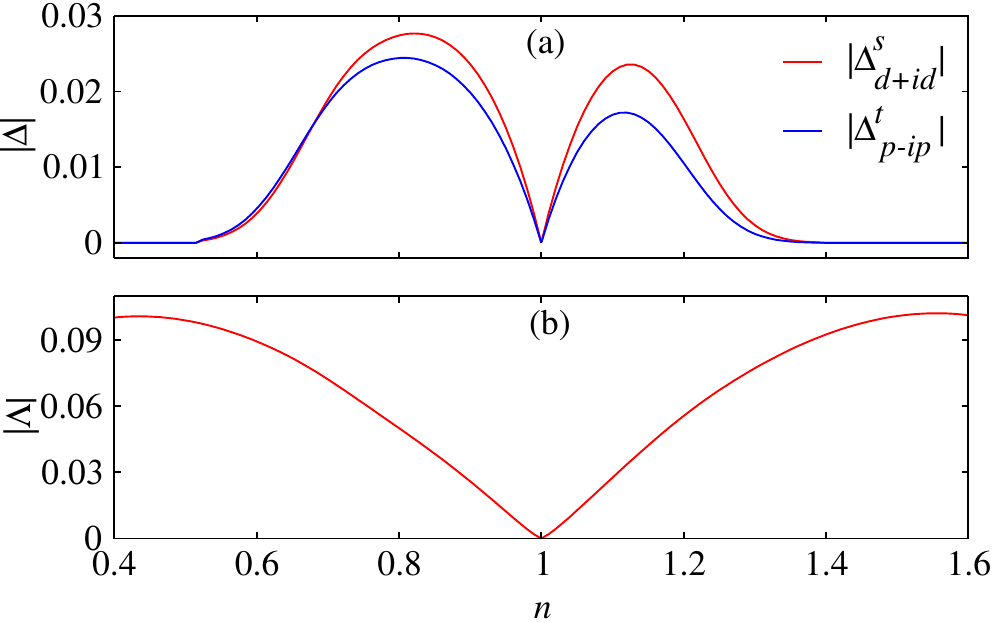}
 \caption{(a) Superconducting gap amplitudes for the spin-singlet $d+id$ and spin-triplet $p-ip$ symmetries  as a function of band filling $n$ for selected values of exchange interaction energy $J=2.62$ meV and displacement field $D=0.4$\;V/nm; (b) The expectation value of the nearest neighbor electron hopping for the same model parameters as in (a).}
 \label{fig:Delta_ndep_J_2_6meV}
\end{figure}

In order to determine the range of realistic values of $J$ we use the formula $J=4|t|^2/U$, where $U$ is the Coulomb repulsion integral originating from the Hubbard model, which serves as a starting point for the $t$-$J$ model derivation. For the case of tWSe$_2$, it has been estimated that $U$ is comparable to the bare band width $W\approx 90$ meV\cite{Wang2020,Haining2020} what would give us $J\approx 3$ meV. Nevetheless, in practice by changing the twist angle one can change the value of $U$, tuning the strength of the correlations. In Fig. \ref{fig:Delta_ndep_J_2_6meV} we show the band filling dependence of the SC gap amplitudes for a selected typical value of $J=2.62$ meV, which corresponds to $U\gtrsim W$ (close to the moderately correlated regime). As can be explicitly seen at half filling ($n=1$) both the pairing and the expectation value of the nearest neighbor electron hopping ($|\Lambda|$) are suppressed, which is a manifestation of the insulating state at half-filling. 

\begin{figure}[!b]
 \centering
 \includegraphics[width=0.45\textwidth]{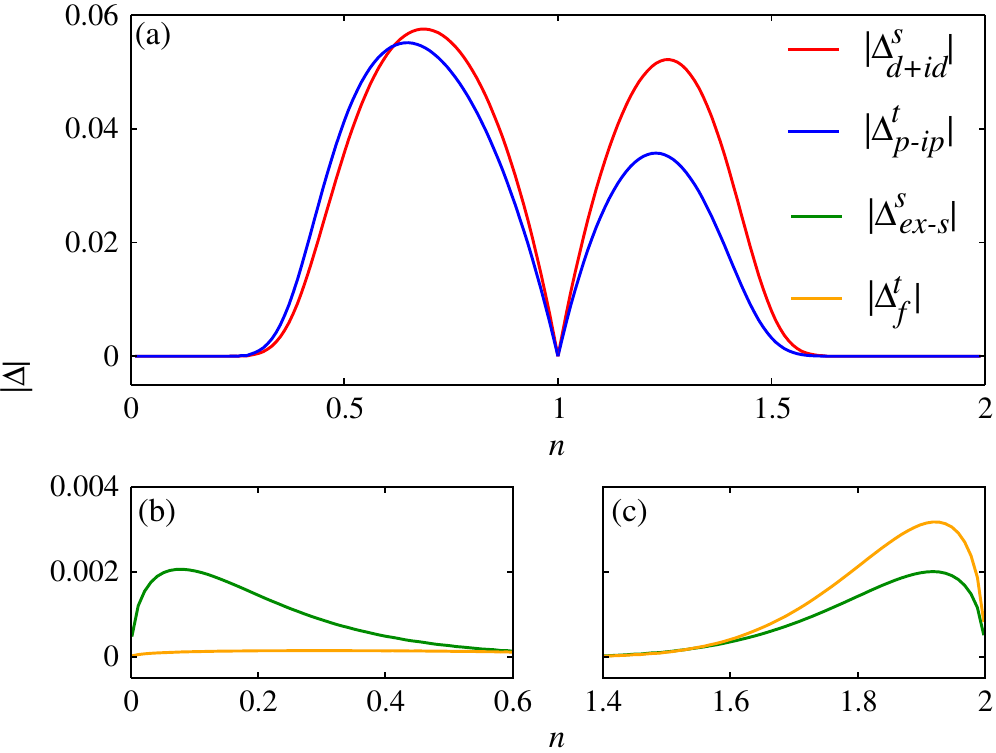}
 \caption{(a) Superconducting gap amplitudes for the spin-singlet $d+id$ and spin-triplet $p-ip$ symmetries  as a function of band filling $n$ for selected values of exchange interaction energy $J=10$ meV and displacement field $D=0.4$; In (b) and (c) we show the low and high electron concentration regime where a mixed singlet $extended$ $s$- and triplet $f$-$wave$ solution becomes stable.}
 \label{fig:Delta_ndep_J_10meV}
\end{figure}

Apart from the $d+id$/$p-ip$-wave solution visible in Figs. \ref{fig:Delt_nJdep} and \ref{fig:Delta_ndep_J_2_6meV} we have also obtained an extended $s$/$f$-wave solution which, however, is much weaker and requires relatively larger values of the $J$ parameter to become stable. In Fig. \ref{fig:Delta_ndep_J_10meV} we provide the results for $J=10$\;meV, where in the low- and high-electron concentration regimes the mentioned extended $s$/$f$-wave paired state is shown to create two additional SC domes as a function of band filling. As one can see, the gap amplitudes $\Delta^s_{ex-s}$ and $\Delta^t_{f}$ are approximately one order of magnitude smaller than the $p$ and $d$-wave solutions $\Delta^s_{d+id}$, $\Delta^t_{p-ip}$. This situation resembles the one obtained earlier for a much simpler model with intersite real space pairing on a square lattice, in which a $d$-$wave$ superconducting state becomes stable around half filling and an extended $s$-wave paired state resides in the low- and high-electron concentration regimes, see Fig. 2 in Ref. \onlinecite{Zegrodnik_Wojcik_2022}. In our current study, due to the triangular lattice and the complex hoppings, the gap symmetry is more exotic being of mixed singlet-triplet type.

\begin{figure}
 \centering 
 \includegraphics[width=0.5\textwidth]{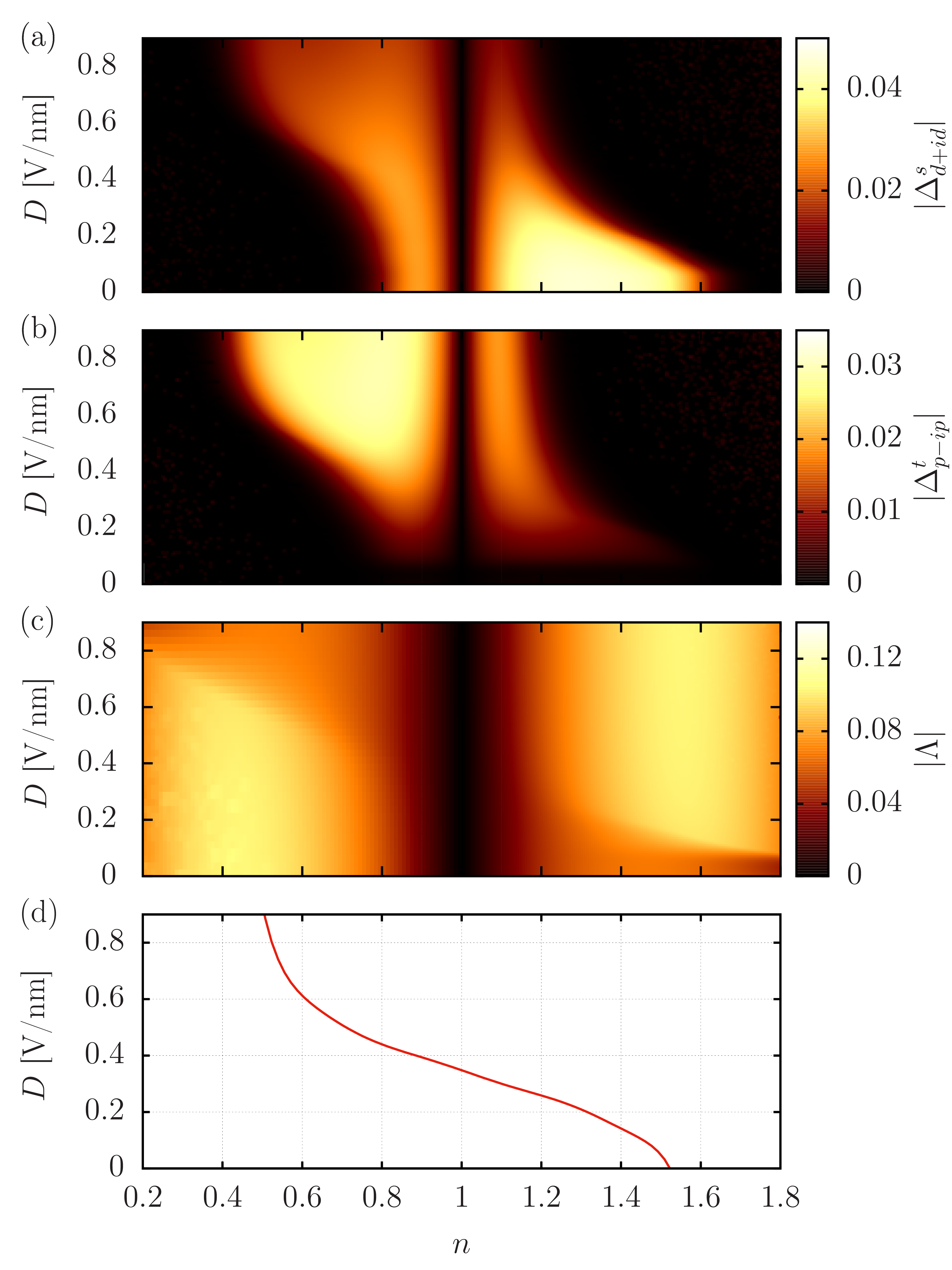}
 \caption{Superconducting gap amplitudes for the spin singlet $d+id$ (a) and spin triplet $p-ip$ (b) symmetries as a function of band filling $n$ and displacement field $D$ for $J=2.62$\;meV. In (c) we show the electron hopping expectation value in the correlated state as a function of $n$ and $D$. The position of the Van Hove singularity which influences the behavior of the superconducting gap is provided in (d).}
 \label{fig:Delta_nDdep}
\end{figure}

Next, we turn to the analysis of the effect of the displacement field on the paired state. For that, we set the typical value of the exchange interaction $J=2.62$ meV, and solve the set of self consistent equations for different complex hopping parameters, which correspond to different displacement fields. In order to make the obtained phase diagram more smooth we have applied interpolation to the data provided Ref.~\onlinecite{Wang2020} where the complex hopping parameters for 14 values of the displacement field in the range $D\in[0,0.9]$\;V/nm have been provided. As one can see in Fig. \ref{fig:Delta_nDdep}, for low values of $D$, the spin-singlet component to the pairing is dominant while for large $D$ the situation changes in favor of the spin-triplet component. It should be noted that for $D\approx 0$ the imaginary part of the complex hoppings is close to zero, meaning that the valley-dependent spin-spitting is very small. This will promote the spin-singlet state. On the other hand, for purely imaginary hoppings one would obtain a pure spin-triplet state (similar to our previous result in Ref. \onlinecite{Zegrodnik2023}). However, this situation is not reached in the considered range of $D$ values. Nevertheless, the result provided in Fig. \ref{fig:Delta_nDdep} shows that by adjusting the displacement field one should be able to tune the balance between the singlet and triplet contributions to the pairing.   

Another effect which is visible in the $(n,D)$-diagram is that the stability regime of the SC state moves towards lower concentrations with increasing $D$. This is caused by the fact that the Van Hove singularity changes its position while increasing the displacement field, see Fig. \ref{fig:Delta_nDdep}(d) and Fig. \ref{fig:FS_and_DOS}(c,d). 
The superconducting state is stabilized by the high density of states at the Van Hove singularity.
Consequently, the behavior of the superconducting gap amplitudes reflects the evolution of the Van Hove singularity with increasing $D$. 
Of course, at half-filling ($n=1$) there is still an insulating state with suppressed pairing amplitudes, regardless of the Van Hove singularity. 
The same applies to the expectation values of the electron hopping which is provided in Fig.~\ref{fig:Delta_nDdep}(c) for completeness. This last result reflects the experimental situation in which also the appearance of the insulating behavior corresponds to the half-filled case in wide range of $D$ values\cite{Wang2020}.

\begin{figure*}
 \centering
 \includegraphics[width=1.0\textwidth]{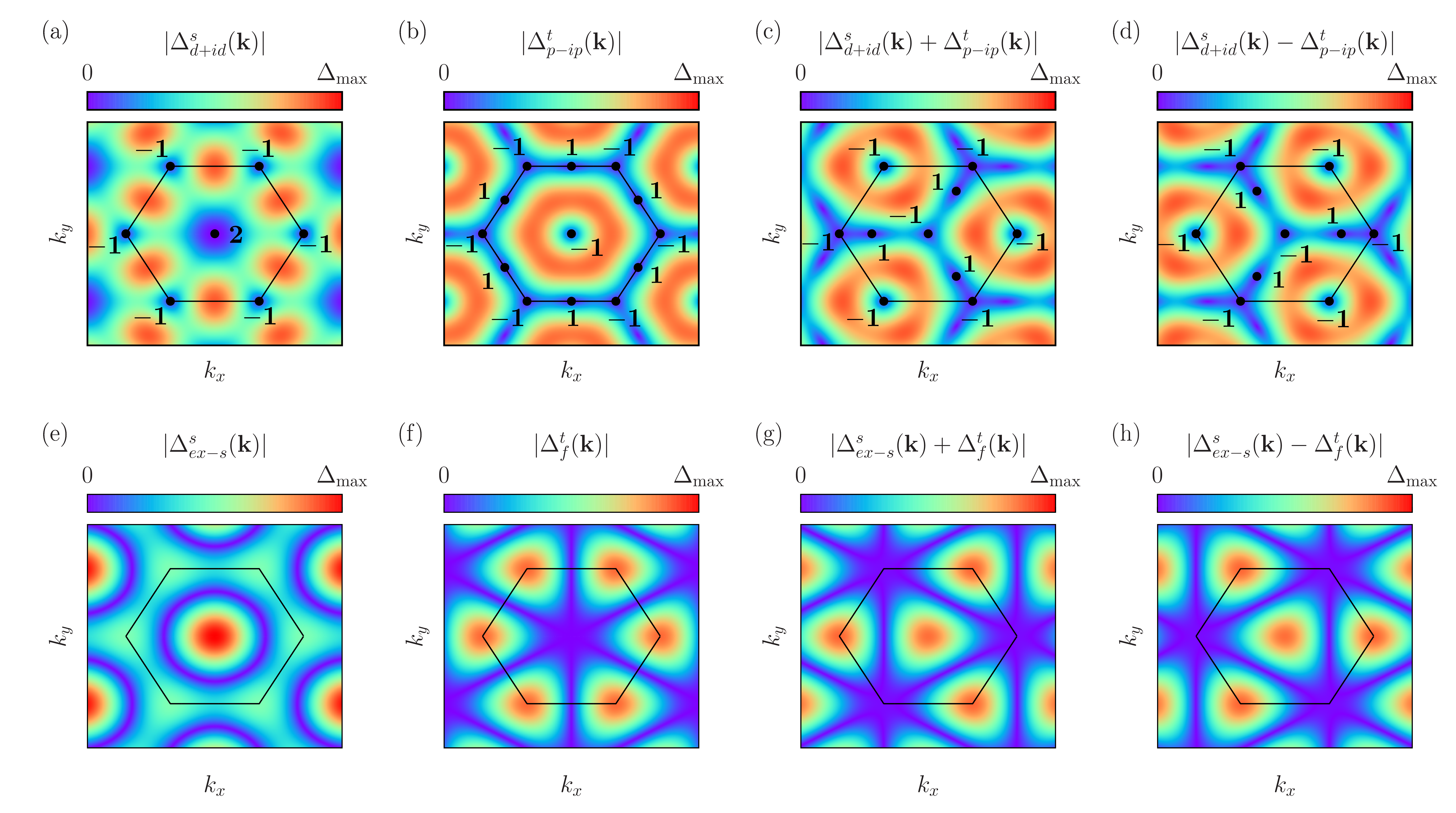}
 \caption{Superconducting gap as a function of momentum for the pure $d+id$-$wave$ (a),  $p-ip$-$wave$ (b), $extended$-$s$-$wave$ (e), and $f$-$wave$ (f) gap symmetries together with the mixed gap symmetries $|\Delta^{s}(\mathbf{k})\pm\Delta^{t}(\mathbf{k})|$ (c,d,g,h). Note that in the mixed states, the $|\Delta^{s}(\mathbf{k})-\Delta^{t}(\mathbf{k})|$ and $|\Delta^{s}(\mathbf{k})+\Delta^{t}(\mathbf{k})|$ gaps open at the spin up and spin down Fermi surfaces, respectively (cf. Eq. \ref{eq:eigenvalues_singlet_triplet}). The nodal points (if they appear) are marked by the black dots together with the topological charge (Chern charge) next to it. The Chern charges provided here have been calculated for the case of electron-like normal state Fermi surface (counterclockwise superconducting gap's phase winding). For the case of hole-like Fermi surfaces the sign of the Chern charges changes to opposite.}
 \label{fig:Delta_kdep_nodal_points}
\end{figure*}

It should be noted that instabilities toward mixed singlet-triplet pairing of $d+id$/$p-ip$ as well as extended $s$/$f$ type has been recently reported in the Hubbard model as applied to the description of tWSe$_2$ in the regime of relatively weak correlations ($U\lesssim 0.7\;W$) within functional renormalization group (FRG) method\cite{Klebl2022}. Also, for very weak Coulomb interactions ($U\lesssim \;0.3W$) a $p+ip$, $d+id$, and $g+ig$ pairing symmetries have been identified and their interplay with spin-polarized pair density waves has been analyzed, however, without the singlet-triplet mixing appearance\cite{Wu_2023_PDW_SC}.

To understand the role of symmetry in the mixed paired state, we provide in Fig. \ref{fig:Delta_kdep_nodal_points} the momentum-dependence of the pure $d+id$-, $p-ip$-, extended-$s$-, and $f$-wave gap symmetries, together with the mixed gap symmetries that have been obtained by us here in proper parameter ranges. As one can see all the pure gap symmetries fulfill the six-fold rotational symmetry, while for the mixed states the symmetry is reduced to $C_3$, which is the one obeyed by our Hamiltonian with valley dependant spin-splitting. Hence, the reduction of symmetry due to the spin-splitting incorporated in the minimal model, naturally leads to singlet-triplet mixing in the superconducting state. Also, it should be noted that for the mixed states, $|\Delta^{s}(\mathbf{k})-\Delta^{t}(\mathbf{k})|$ and $|\Delta^{s}(\mathbf{k})+\Delta^{t}(\mathbf{k})|$, gaps open at the spin up and spin down Fermi surfaces, respectively (cf. Eq. \ref{eq:eigenvalues_singlet_triplet}). 

As can be seen from Fig. \ref{fig:Delta_kdep_nodal_points} the $d+id$ and $p-ip$ gap symmetries as well as their mixtures all have nodal points in the Brillouin zone for which the gap is exactly zero (marked by black dots). This is not the case for the extended-$s$-, and $f$-wave gaps, as well as their mixtures, for which only nodal lines appear. Each nodal point comes with a topological charge (Chern charge) which is also provided in the figure and can lead to a nonzero Chern number of a given paired state. A detailed analysis of the topological features of the mixed $d+id/p-ip$ superconducting state is presented in the next subsection.


\begin{figure}
 \centering
 \includegraphics[width=0.5\textwidth]{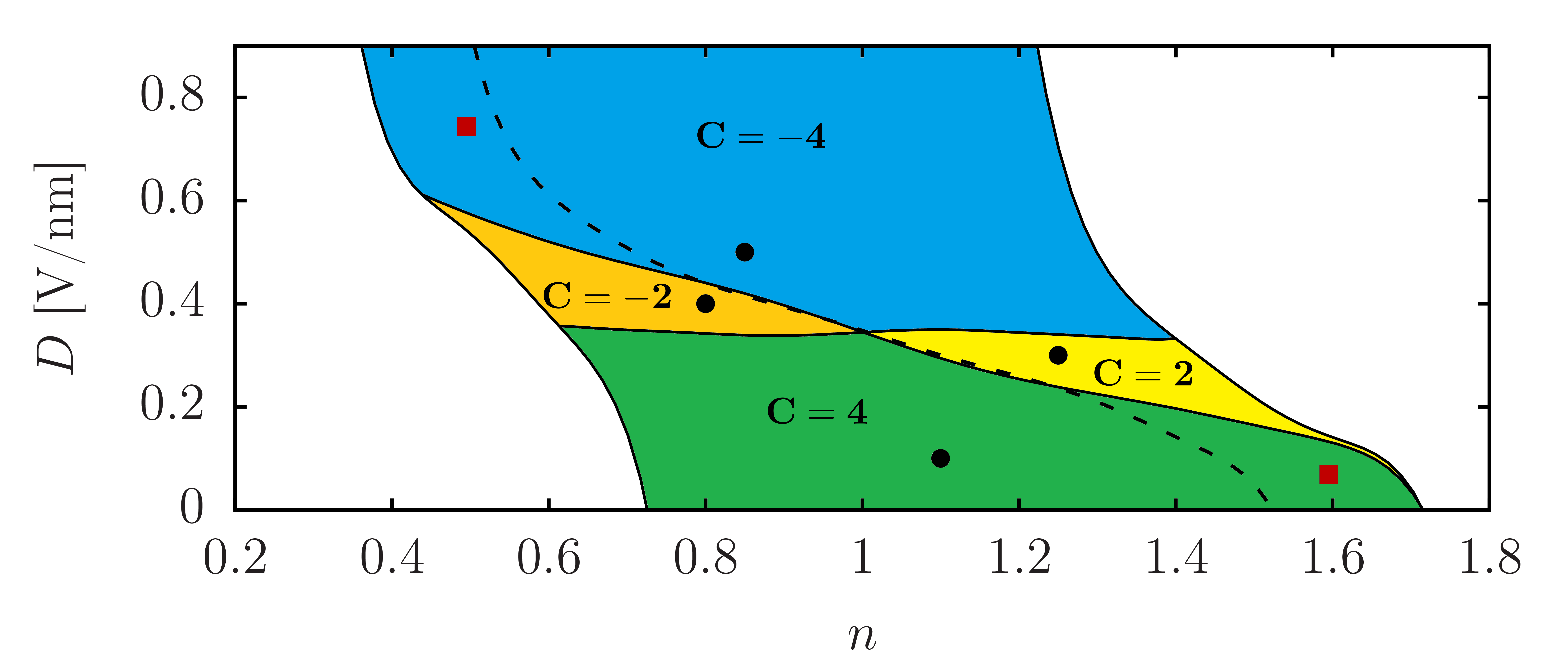}
 \caption{The topological phase diagram of the mixed $d+id$ (singlet)/$p-ip$ (triplet) superconducting state in the $(n,D)$ plane. The colored regions correspond to the stability of the mixed paired state (cf. Fig. \ref{fig:Delta_nDdep}). Different colours correspond to different values of the Chern number. The position of the Van Hove singularity in the diagram is marked by the dashed line. By changing the experimentally controllable parameters, like electron concentration ($n$) and displacement field ($D$), one can induce topological phase transitions. The black dots correspond to $n$ and $D$ values which have been selected for detailed analysis provided in Fig. \ref{fig:Delta_kdep_nodal_points2}. Additionally, the red squares mark the two selected situation of relatively high and low $n$ values which are analyzed in detail in Fig. \ref{fig:Delta_kdep_nodal_points_additional}.
 }
 \label{fig:nDdep_topo}
\end{figure}

\begin{figure*}
 \centering
 \includegraphics[width=1.0\textwidth]{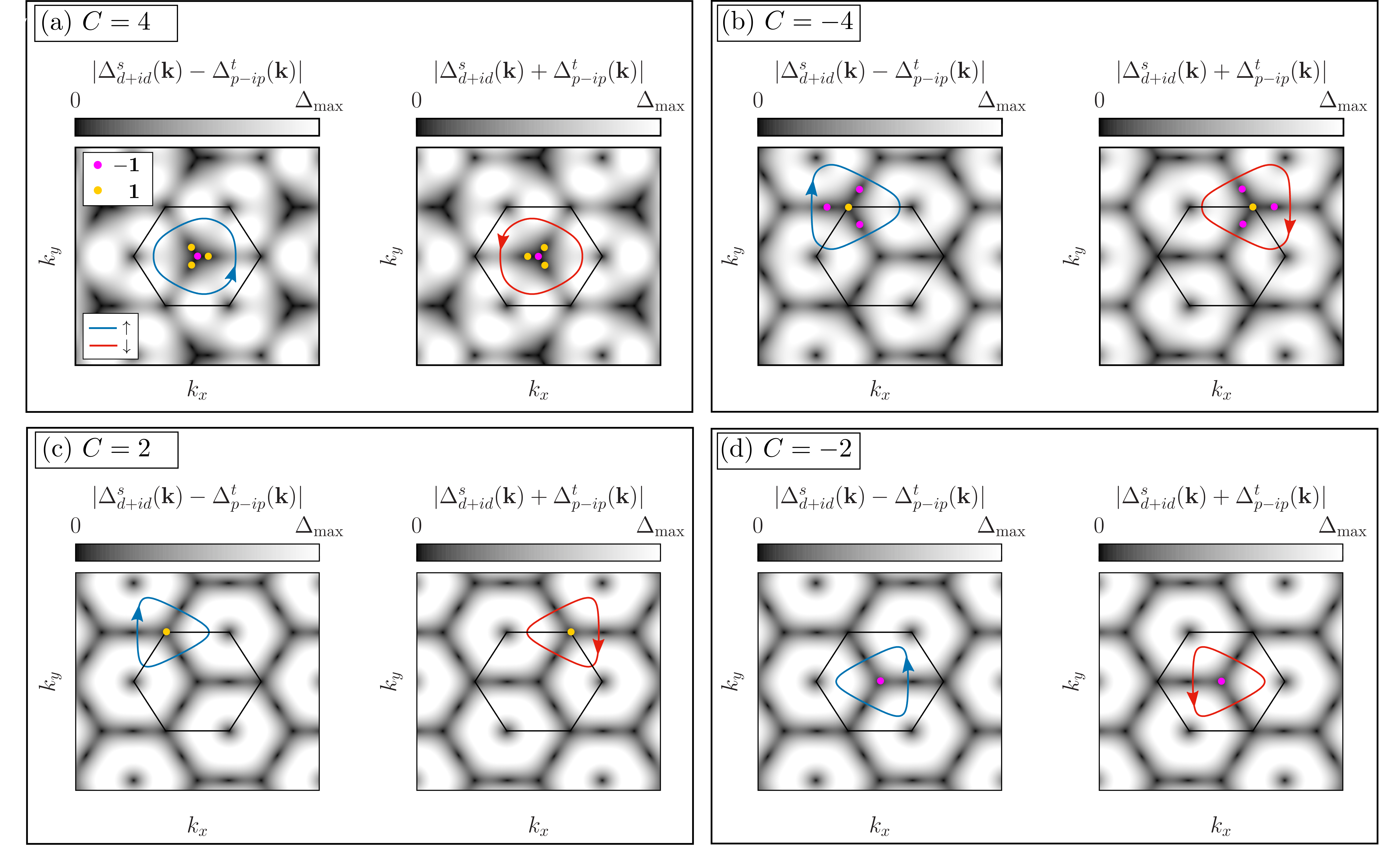}
 \caption{The normal state Fermi surfaces for spin up and down electrons as well as the $\mathbf{k}$-dependent SC gaps for four selected values of $n$ and $D$, which are marked by black dots in Fig. \ref{fig:nDdep_topo} and correspond to following values of the Chern number: $C=4$ (a), $C=-4$ (b), $C=2$ (c), and $C=-2$ (d). Note, that the value of the Chern number is simply the sum of the Chern charges contained inside the spin up and down Fermi surfaces (marked by the colored dots). The arrows at the Fermi surfaces represent the winding direction which is counterclockwise (clockwise) for the electron-like (hole-like) Fermi surfaces. Note that the signs of the Chern charges depend on the winding direction.}
 \label{fig:Delta_kdep_nodal_points2}
\end{figure*}

\begin{figure*}
 \centering
 \includegraphics[width=1.0\textwidth]{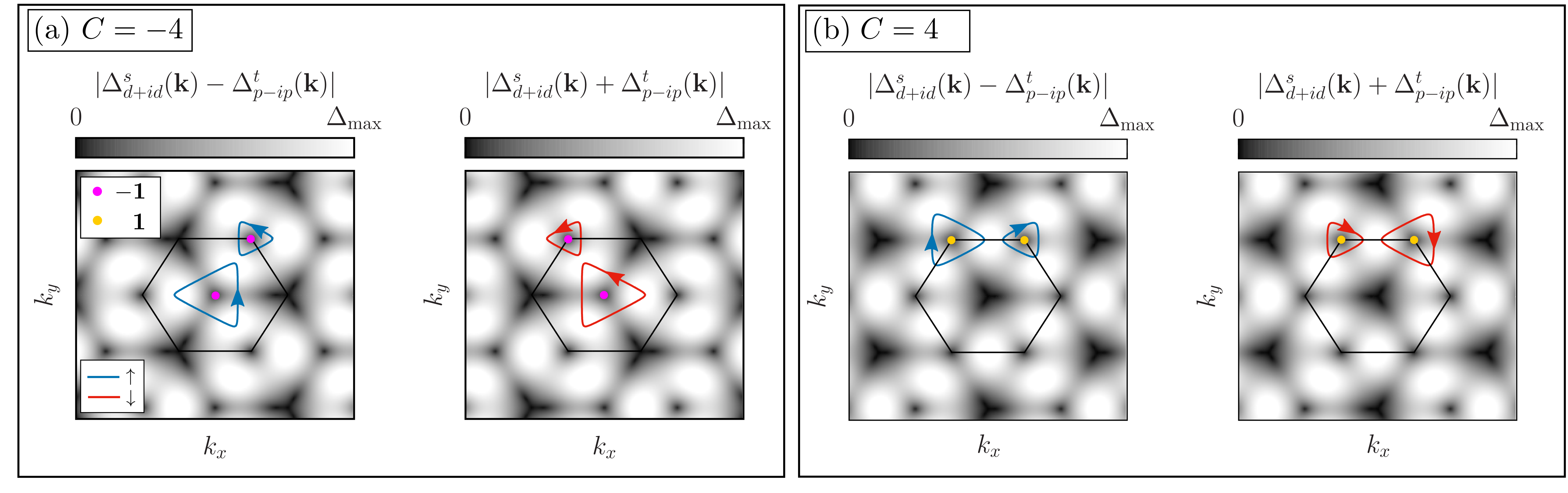}
 \caption{The same as in Fig. \ref{fig:Delta_kdep_nodal_points2} but for $n$ and $D$ values which are represented by the red squares marked in Fig. \ref{fig:nDdep_topo} and correspond to following values of the Chern number: $C=-4$ (a) and $C=4$ (b). In contrast to Fig. \ref{fig:Delta_kdep_nodal_points2}, now we have two Fermi surfaces per spin. Note, that in the low- and high-$D$ regime crossing the Van Hove singularity line and the resulting transition from electron- to hole-like Fermi surface no longer leads to topological phase transition.}
 \label{fig:Delta_kdep_nodal_points_additional}
\end{figure*}

\subsection{Topological properties of the paired state}
\label{Sec:ResultsB}
We now analyze in detail the topological properties of the mixed singlet-triplet superconducting state. Since the paired states studied here spontaneously break time-reversal symmetry, their classification is given by the Chern number\cite{Qi.2011}. We calculate the Chern number with the use of the Brillouin zone triangulation method\cite{Hiroshi2005_chern_number}. As mentioned in the previous section, the extended-$s$/$f$-wave paired state does not have any nodal points in the Brillouin zone, and is therefore topologially trivial with $C=0$. On the other hand, the calculated Chern number for the $d+id/p-ip$ symmetry takes the values of $C=\pm 2,\;\pm 4$, depending on both band filling and displacement field. In Fig.~\ref{fig:nDdep_topo} we show the topological phase diagram of the mixed $d+id/p-ip$ paired sate in the $(n,D)$-plane. As can be seen according to our calculations, by changing the experimentally controllable parameters one can induce topological phase transitions between regimes characterized by different values of the Chern number. It should be noted that the bare band of the considered model is topologically trivial, therefore, the topological features are introduced by the paired state itself.

While analysing the data provided in Fig. \ref{fig:nDdep_topo} it is worth emphasizing that the Chern number can be interpreted as the winding number of the superconducting gap's complex phase while going around the normal state Fermi surface (FS):\cite{Pavarini:884084}
\begin{equation}
    C=\frac{1}{2\pi}\oint_{FS}d\mathbf{l}\cdot \nabla\phi,
    \label{eq:winding_number}
\end{equation}
where for the case of electron-like (hole-like) FS the integration is carried out counterclockwise (clockwise). Therefore, the shape and size of the FS, which can be tuned by both $n$ and $D$, may affect the resulting topological features of the SC state. In fact, Eq.~\eqref{eq:winding_number} can be further simplified, as the Chern number simply counts the Chern charge of the nodal points which are contained within the FS,
\begin{equation}
    C=\sum_{i\in \Omega_{FS}}(-1)^wC_i,
    \label{eq:Chern_partial}
\end{equation}
where $C_i$ is the Chern charge of a given nodal point which can be determined by calculating the winding number of the order parameter around it. The sign of the Chern charge depends on whether it is enclosed by an electron- ($w=0$) or hole-like ($w=1$) Fermi surface. The $i$ index runs only over the nodal points that are contained inside the Fermi surface. In Fig. \ref{fig:Delta_kdep_nodal_points} we provide the Chern charges, $C_i$, of all the nodal points for the considered gap symmetries. Once we know the Chern charge of every nodal point, as well as the shape of the FS, we can determine the Chern number (winding number) by simply using Eq. (\ref{eq:Chern_partial}). With this procedure one can easily interpret the obtained topological phase diagram by looking at the relative position of the Fermi surface and the Chern charges of the nodal points. 

It should be noted that the nodal points of the $d+id/p-ip$ paired state fall into two categories. The first one corresponds to the nodes located at the high symmetry points ($\Gamma$, $K$, and $K'$) which are fixed and result from the symmetry of the triangular lattice itself. The second category are the points which position in the Brillouin zone depends on the balance between the singlet and triplet component to the pairing. There are three such points per each of the two mixed gaps ($|\Delta^{s}(\mathbf{k})-\Delta^{t}(\mathbf{k})|$ and $|\Delta^{s}(\mathbf{k})+\Delta^{t}(\mathbf{k})|$) and they lie at the $\Gamma$-$K$ and $\Gamma$-$K'$ connecting lines. The movement of the mentioned nodal points inside the Brillouin zone, induced by tuning $n$ and $D$ can affect the topological features as we show in some more detail further on.

In Fig.~\ref{fig:Delta_kdep_nodal_points2} we show the normal state Fermi surfaces as well as the momentum-dependent SC gap for four representative points that are marked by black dots in the phase diagram (Fig. \ref{fig:nDdep_topo}) and are characterized by different values of the Chern number. The spin up and spin down Fermi surfaces together with the corresponding SC gaps (cf. Eq. \ref{eq:eigenvalues_singlet_triplet}) are provided in separate Figures in the panel. As one can see, the topological transitions that appear across the phase diagram provided in Fig. \ref{fig:nDdep_topo} are not induced by a change in order parameter symmetry. But rather, they result from the fact that with changing $n$ and $D$ one influences the normal state Fermi surface as well as one moves the nodal points that lie in between the $\Gamma$-$K$ and $\Gamma$-$K'$ points. This in turn influences the total Chern charge contained inside the Fermi surfaces. In particular, the transitions from $C=4$ (a) to $C=-2$ (d) as well as from $C=-4$ (b) to $C=2$ (c), both correspond to moving the three Chern charges laying at the $\Gamma$-$K$ and $\Gamma$-$K'$ connecting lines outside the interior of the Fermi surfaces. What is also important is that with changing $n$ and $D$ one can induce transitions from electron-like to hole-like Fermi surface. This changes the position of the Fermi surface as well as the signs of the Chern charges what influences the overall Chern number. That is why the transition line between $C=4$ and $C=2$ as well as $C=-4$ and $C=-2$ coincides  with the Van Hove singularity line in the regime of moderate displacement fields ($D\sim 0.3-0.5$\;V/nm).

However, outside the region of moderate displacement fields the transition between electron-like and hole-like Fermi surface no longer induces a topological phase transition. This is caused by the fact that now at the Van Hove singularity a transition between one FS per spin to two FS per spin appears. This prevents from changing the value of the Chern number. In Fig. \ref{fig:Delta_kdep_nodal_points_additional} we show two such situations with two normal state Fermi surfaces per spin.
For example, in the low-$D$ regime, crossing the Van Hove singularity leads to transition from a single electron-like FS centered at the $\Gamma$ [Fig. \ref{fig:Delta_kdep_nodal_points2} (a)] to two hole-like FS centered at the $K$ and $K'$ [Fig. \ref{fig:Delta_kdep_nodal_points_additional} (b)]. Nevertheless, as one can see it does not lead to a change of the total Chern charge contained inside the Fermi surfaces. As a consequence of this effect, the topological transition line in Fig. \ref{fig:nDdep_topo} does not coincide with the Van Hove singularity line at its full extent. 


\section{Conclusions}
\label{Sec:Outlook}
We have applied the $t$-$J$ model to the description of the superconducting state in tWSe$_2$ within the Gutzwiller approach. As we show, such an approach reproduces the experimental situation with two superconducting domes as a function of band filling, and a correlation-induced insulating state at half-filling\cite{Wang2020}. Consistent with the experimental data, the insulating state is not affected by the displacement field in wide range of $D$. According to our calculations the superconducting domes have a mixed $d+id$ (singlet) and $p-ip$ (triplet) symmetry, which is a direct consequence of the valley-dependent spin splitting. If such an exotic paired state is indeed realized in this system, it would possible to tune the balance between the singlet and triplet contributions to the pairing with the use of the displacement field (cf. Fig. \ref{fig:Delta_nDdep}). Namely, for low values of $D$, the singlet pairing dominates and for high values of $D$ the triplet pairing takes over. Additionally, due to the evolution of the density of states, with increasing $D$, the paired state stability regime moves towards lower electron concentrations. Furthermore, our model reveals a weak extended $s$-wave and $f$-wave mixed paired state for relatively large values of the $J$, at low- and high-electron concentrations.

According to our analysis the mixed $d+id$ (singlet) and $p-ip$ (triplet) paired state is topologically non-trivial with Chern numbers $C=\pm2,\;\pm4$, depending on the value of band filling and displacement field (see Fig. \ref{fig:nDdep_topo}). The changes of the Chern number are caused by: (i) changes of the normal state Fermi surfaces: either due to a change in size or due to a Lifshitz transition from an electron-like to a hole-like Fermi surface (cf. Fig. \ref{fig:Delta_kdep_nodal_points}); (ii) The movement of the nodal points located at the $\Gamma$-$K$ and $\Gamma$-$K'$ connecting lines. By tuning the displacement field and electron density one could, according to our results, induce topological phase transitions between pairing states characterized by different values of the Chern number.

Our methods find the zero temperature ground state of a relevant $t$-$J$ model for twisted WSe$_2$. We can estimate the critical temperature by looking at the maximum magnitude of the gap, $|\tilde{\Delta}|_{\rm max} \approx 0.06J$. This suggests a critical temperature of the order of $T_c \approx 1$ K, which is at a temperature slightly below the lowest temperature of the recent transport experiments\cite{Wang2020}. This might explain why so far only `signatures' of superconductivity have been observed. It is therefore important to perform these transport measurements at lower temperatures, to verify the existence of superconductivity. To experimentally detect the nontrivial nature of the pairing state, we suggest  Knight shift measurements to detect the non-singlet component. A detection of the Kerr effect can reveal the spontaneous broken time-reversal symmetry. Finally, our prediction for topological superconductivity in tWSe$_2$ suggests this system is ideal for observing nontrivial chiral edge states and possible Majorana particles.\cite{Qi.2011}

The  code  which  was  written  to  carry  out  the  numerical calculations as well as the data behind the figures are available in the open repository\cite{zegrodnik_michal_2024_zenodo}.

\begin{acknowledgments}
We acknowledge discussions with Mark Fischer.
This research was funded by National Science Centre, Poland (NCN) according to decision 2021/42/E/ST3/00128. L.R. is supported by the Swiss National Science Foundation through Starting Grant No. TMSGI2\_211296. For the purpose of Open Access,
the author has applied a CC-BY public copyright licence to any Author Accepted Manuscript (AAM) version arising from this submission.
\end{acknowledgments}

\bibliography{refs.bib}

\end{document}